\documentclass[draftcls,onecolumn,12pt]{IEEEtran}
\usepackage{amsmath,amsfonts}
\usepackage{array}
\usepackage{url}
\usepackage{cite}
\usepackage{url}
\usepackage{booktabs}
\usepackage{braket} 
\usepackage{graphicx}
\usepackage{algorithm}
\usepackage{algpseudocode}
\usepackage{siunitx}

\usepackage{mathtools}

\newcommand{\mytitle}{Hyperdimensional Computing for ADHD Classification using EEG Signals}

\begin{document}

\title{\mytitle}

\author{Federica~Colonnese, Antonello~Rosato, Francesco~Di~Luzio, and Massimo~Panella
\thanks{All authors are with the Department of Information Engineering, Electronics and Telecommunications, University of Rome ``La Sapienza'', 00184 Rome, Italy (e-mail: andrea.ceschini@uniroma1.it, massimo.panella@uniroma1.it).}
\thanks{Corresponding author: Prof. M. Panella, Via Eudossiana 18, 00184 Rome, Italy (phone: +39-0644585496; e-mail: massimo.panella@uniroma1.it).}
}

\maketitle

\begin{abstract}
Following the recent interest in applying the Hyperdimensional Computing paradigm in medical context to power up the performance of general machine learning applied to biomedical data, this study represents the first attempt at employing such techniques to solve the problem of classification of Attention Deficit Hyperactivity Disorder using electroencephalogram signals.
Making use of a spatio-temporal encoder, and leveraging the properties of HDC, the proposed model achieves an accuracy of $88.9\%$, outperforming traditional Deep Neural Networks benchmark models.
The core of this research is not only to enhance the classification accuracy of the model but also to explore its efficiency in terms of the required training data: a critical finding of the study is the identification of the minimum number of patients needed in the training set to achieve a sufficient level of accuracy. 
To this end, the accuracy of our model trained with only $7$ of the $79$ patients is comparable to the one from benchmarks trained on the full dataset.
This finding underscores the model's efficiency and its potential for quick and precise ADHD diagnosis in medical settings where large datasets are typically unattainable.
\end{abstract}


\section{Introduction}
\label{sect:introduction}
Hyperdimensional Computing (HDC) techniques, also called Vector Symbolic Architectures (VSA), have gained increasing attention in the fields of Artificial Intelligence (AI) and neural sciences. 
This brain-inspired computational approach has been widely applied in a growing number of domains, including signal recognition \cite{imani2017voicehd}, biomedical and image classification \cite{burrello2019laelaps, ge2020classification}.
It draws inspiration from neuroscience research \cite{kanerva2009hyperdimensional}, and has showcased significant promises in tackling intricate challenges related to data representation and processing, which nowadays are areas largely ruled by Deep Neural Networks (DNNs).
DNNs encounter significant challenges as their computational expenses can be heavy, especially for complex problems: the exponential increase in parameters leads to higher computational complexity and memory requirements, prolonged training times due to the optimization of numerous parameters, and the need for large datasets.

The scientific community is exploring integrating HDC principles with DNN frameworks to address, for example, complexity and data requirements. 
HDC involves projecting data onto thousands or millions of dimensions, moving away from traditional numerical representations, and fundamentally changing the way information is handled \cite{kanerva2009hyperdimensional}.
As a result, HDC techniques are increasingly employed in a multitude of domains, ranging from the biomedical field \cite{schindler2021primer}, to the identification of driving styles \cite{chen2023driving}, as well as speech recognition \cite{imani2017voicehd}, highlighting the adaptability and significance of HDC methodologies across various sectors.
HDC is attracting interest for its potential in scenarios needing alternatives to current techniques, particularly due to its ability to handle complex data situations effectively.

This paper examines an application of HDC to the biomedical domain, specifically to the detection of Attention Deficit Hyperactivity Disorder (ADHD) using Electroencephalogram (EEG) signals, recorded while patients were resting with open eyes.
To this end, ADHD, one of the most prevalent neurodevelopmental disorders, impacts $4\%-12\%$ of young children and $4\%-5\%$ of adults \cite{wilens2010understanding}. 
According to the criteria established by the Diagnostic and Statistical Manual of Mental Disorders (DSM-5) \cite{american2013diagnostic}, diagnosing ADHD generally depends on subjective methods. These include behavioral assessments to evaluate how individuals socialize, behave, and communicate, along with standardized diagnostic tools such as questionnaires completed by parents and teachers to observe behavior across different settings.
However, ADHD's misdiagnosis is common due to symptoms overlapping with other conditions and neurodevelopmental disorders itself, insufficient training among healthcare professionals, and the stigma associated with this condition leading also to under-reporting, posing challenges for accurate diagnosis procedures \cite{hartnett2004gifted}.

To the best of our knowledge, this study represents the first attempt at integrating HDC with ADHD classification.
On this end, the key distinction and main novelty aspects lie, not only in achieving higher accuracy rates compared to DNNs but, mostly in accomplishing this with a reduced amount of patient data inputs: by training the proposed framework using EEG signals recorded on a very small subset of patients (seven patients, meaning the $8.86\%$ of the dataset), it was possible to obtain a classification accuracy of $72.9\%$, already surpassing the benchmark DNNs trained on the whole dataset.
These improvements led to a quick, efficient tool that could merge DSM-5 methods with automation, speeding up diagnoses, enhancing accuracy, and reducing the workload for healthcare professionals by simplifying processes and decreasing data preparation.

The code developed in this research will be made available upon request.

\section{Current Literature and Main Goals}
This section briefly reviews current HDC applications for neurodevelopmental disorders and introduces the novel contributions proposed by our approach in this field.

\subsection{State of The Art} 
\label{sect:sota}
In the biomedical domain, HDC classification methods have demonstrated exceptional performances across different applications, including but not limited to seizure detection \cite{burrello2019hyperdimensional}, gesture recognition \cite{rahimi2016hyperdimensional}, EEG error-related potentials classification for brain-computer interfaces \cite{rahimi2017hyperdimensional}, and emotion recognition \cite{menon2022efficient}. 
Furthermore, it has exhibited remarkable classification accuracy even when confronted with challenges such as limited training data \cite{chang2019hyperdimensional}, one-shot learning situations \cite{rosato2022few}, and frequent model updates. These characteristics present significant advantages for medical physiological sensors tailored to user-specific data, given the susceptibility of these sensors to displacements between runs or during extended periods of use \cite{gorjan2022removal}.

The growing interest in biomedical research for ADHD detection is largely driven by advancements in medical technology. 
Imaging techniques are frequently employed in this field and DNNs are the primary methods for analyzing these complex datasets due to their ability to identify intricate patterns as, for instance, in brain activity from functional magnetic resonance imaging (fMRI) \cite{riaz2020deepfmri,zou20173d,colonnese2023fast,tian2008enhanced} and EEG \cite{mohammadi2016eeg,vahid2019deep}, as well as in extended behavioral analysis and emotion recognition from multivariate data and time series \cite{liparulo2017,DILUZIO2023104418}.
However, DNNs encounter challenges such as limited dataset availability and the need for complex, resource-intensive networks to effectively process those large-scale medical data.
Despite these advancements, to the best of our knowledge, there is currently no known application of HDC for the classification of ADHD or neurodevelopmental disorders in general. 
As will be shown in the following paragraphs, HDC could offer potential benefits in this area due to its efficiency and robustness in handling high-dimensional data. 
Exploring HDC might lead to new, less resource-intensive methods for diagnosing and understanding ADHD, opening new opportunities for research and application in neurodevelopmental studies.

Recent advancements in the classification of ADHD, have leveraged the potential of Transfer Learning (TL) and pre-trained Convolutional Neural Network (CNN) models \cite{sadiq2022exploiting}. These approaches offer a robust framework for overcoming challenges associated with high-dimensional data, variances in data acquisition, and class distribution imbalances. Specifically, Transfer Learning has shown significant promise in scenarios with limited labeled data by generalizing patterns learned from large datasets to task-specific applications. For instance, \cite{zhang2019use} demonstrated the effectiveness of TL using the ResNet-50 architecture to classify ADHD from functional Magnetic Resonance Imaging (fMRI) data, achieving a classification accuracy of $93.45\%$. Moreover, the Class Activation Map (CAM) analysis revealed distinct differences in brain activity between children with ADHD and healthy controls, highlighting critical areas such as the frontal, parietal, and temporal lobes. 

Beyond deep learning methods, in \cite{sato2012evaluation} the predictive power of three feature extraction methods (regional homogeneity - ReHo), amplitude of low-frequency fluctuations - ALFF), and independent component analysis maps) is evaluated, combined with 10 pattern recognition algorithms for ADHD classification. Their results showed that while ALFF and ReHo maps provided useful discriminatory information for differentiating ADHD patients from typically developing (TD) controls, accuracy remained limited. 
However, the inclusion of RSN features significantly improved performance in distinguishing combined vs. inattentive ADHD subtypes.
The study emphasized that relevant discriminative information is spatially distributed across the entire brain rather than confined to specific regions. 

Similarly, advancements in EEG-based classification were demonstrated by \cite{tor2021automated}, who developed a novel automated system (AS) to classify ADHD. Their method employed empirical mode decomposition (EMD), discrete wavelet transform (DWT), and autoregressive modeling coefficients, combined with adaptive synthetic sampling (ADASYN) to balance the dataset. The highest accuracy, $97.88\%$, was achieved using the K-Nearest Neighbor (KNN) classifier on EEG data from 123 children. This study highlights the potential of automated classification systems as supplementary diagnostic tools, especially when combined with advanced feature extraction techniques and robust sampling strategies.

In general, EEG has become a powerful and non-invasive technique and is nowadays used in different domains, ranging from improving meditation techniques \cite{khadam2024enhancing} to detecting ADHD. The ongoing research focuses on the refinement of signal processing methods and deep learning models to maximize diagnostic precision. The integration of advanced machine learning approaches, particularly CNN and transfer learning, continues to push the boundaries of EEG-based diagnosis, offering new possibilities for earlier and more reliable detection of ADHD.

The field of EEG-based ADHD detection has made considerable progress through the application of advanced signal processing and machine learning techniques aimed at improving diagnostic accuracy and robustness. Among the most effective methods are signal decomposition techniques, which have shown potential in extracting key features from EEG signals that are critical for classification tasks \cite{Akbari2021, Akbari2023}. Techniques such as empirical mode decomposition (EMD) and wavelet transforms, commonly employed in motor imagery classification, have demonstrated promising results\cite{sadiq2019motor}.
In recent years, CNNs have revolutionized the way EEG signals are processed and analyzed; they have proven to be highly effective for decoding complex patterns in raw EEG data, enabling the identification of subtle changes in brain activity without the need of preprocessing them \cite{schirrmeister2017deep}.

\subsection{Novelty of The Proposed Approach}
\label{subsect:novelty}
The primary distinction and innovative aspect of this work is the ability to achieve higher accuracy rates compared to well-known DNNs while using a smaller amount of patient data.
The main goal of our paper is twofold: firstly, we want to classify ADHD more rapidly using HDC techniques: this effort aims at creating rapid and lightweight tools that can seamlessly fit into real-world clinical settings. 
Combining formal diagnostic methods from the DSM-5 \cite{american2013diagnostic} with automated tools could be crucial for medical professionals since it could help in reducing diagnosis time and workload. 
The second goal focuses on working with EEG data that has only been subjected to standard signal processing steps, without any specialized medical preprocessing: this allows to potentially avoid the need for specific medical preprocessing techniques, streamlining the workflow, and removing the complexities and intrinsic unreliability of standard, human-based preprocessing.
This approach not only simplifies the overall process but also alleviates the burden on medical professionals, enabling a more direct focus on diagnostic aspects rather than dedicating significant efforts to data preparation.
This approach is driven by the understanding that, despite considerable progress in research, clinical knowledge of these disorders remains in the early stages since the exact distinctions and changes that define them have not been fully determined.
Their complexity is rooted in the elaborate nature of brain functioning and specific challenges associated with neurodevelopmental disorders. 
On top of that, medical data is also subject to standard privacy-preserving practices that hinder the processability of the data.

Regarding the specifics of our method, the increased accuracy of our spatio-temporal HDC encoder is inherently linked to the encoding of data in a hyperdimensional space, where the properties of high dimensionality and uniform distribution of vector spaces enable the isolation and emphasis of significant patterns.
Another advantage of our study is the training efficiency achieved by HDC: our approach showcases significantly reduced training times: this quality of HDC accentuates its efficiency, making it an ideal solution for environments with limited storage capabilities. 
This is particularly relevant in healthcare applications, where HDC can be utilized in clinical settings that may not have the capacity for high-powered computing systems. 
Lastly, HDC's low demand for GPU resources makes it well-suited for medical environments, ensuring that effective and efficient data processing can be achieved. 

\section{Fundamentals of Hyperdimensional Computing}
\label{sect:hdc_fundamentals}
\begin{figure*}[!ht]
    \centering
    \includegraphics[width=0.9\textwidth]{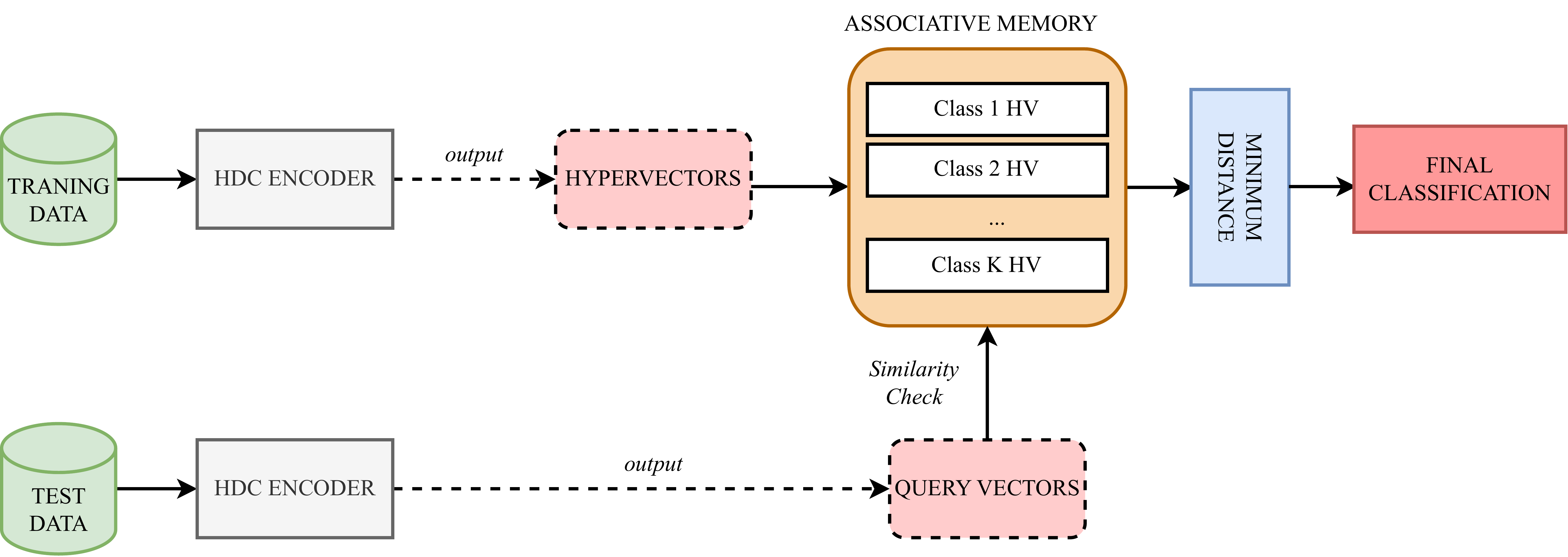}
    \caption{General scheme of a general HDC classification problem.}
    \label{fig:general_hdc_scheme}
\end{figure*}

%
Our approach is based on the foundational scheme of HDC's theoretical background, and it is primarily inspired by the framework introduced in the paper by Abbas Rahimi \emph{et. al} \cite{rahimi2017hyperdimensional} while incorporating specific modifications to suit new objectives and purposes.
A general scheme of the HDC framework is presented in Fig.~\ref{fig:general_hdc_scheme} and it will be discussed in the following. 
This section will provide a concise overview of the essential ideas and components that define HDC as a complete computational architecture.

\subsection{Hypervectors and Main Operations}
\label{sect:HV_operations}
HDC relies on ultra-wide elements known as hypervectors (HVs), which typically have dimensionality in the tens of thousands \cite{kanerva2009hyperdimensional}. Unlike traditional localized representations, information in HDC is distributed across all components of these vectors, making the system inherently robust \cite{kleyko2022vector}: one of the core advantages of high-dimensional (HD) representations is their capacity to tolerate small variations between similar patterns, with this tolerance increasing as the dimensionality of the HVs grows \cite{kanerva2009hyperdimensional}.

HDC often employ random vectors to leverage the concentration of measure phenomenon, which states that in high-dimensional spaces, random vectors are nearly orthogonal, allowing distinct objects to be represented with almost perpendicular vectors \cite{ledoux2001concentration} which is a crucial property for effectively distinguishing between different pieces of information.

Given that HVs serve as the fundamental computational building blocks for constructing VSAs, it is essential to establish precise properties for operations on HVs that can enable the use of VSAs as a comprehensive computational framework. These operations include binding, which combines two HV to represent relationships between concepts or features; bundling, which aggregates multiple HVs into a composite representation that captures the essence of the combined data; and permutation, which rearranges the elements of a HV, ensuring that the temporal order of information is maintained.
They allow for the manipulation and association of data through algebraic operations in ways that can vary depending on the nature of the hyperdimensional vectors used and the information they contain. HDC employs operations that vary based on objectives and data structure, leading to fast learning, high energy efficiency, and accuracy in learning and classification tasks \cite{kanerva2009hyperdimensional}.

Also, the high dimensionality of HVs offers significant advantages for information separation and distinction: since data is distributed across all components of an HV, no single component holds the entire representation. 
This distributed nature makes HDC robust to noise, as altering a few components of the HV does not significantly impact the overall data representation, maintaining the system's integrity.

HVs in HD computing can be diverse (binary, real, bipolar, etc.) \cite{schlegel2022comparison}, influencing the processing function depending on the computational architecture. 
Similarity metrics, such as Cosine Similarity and Hamming Distance, are used for measuring the correlation between HVs \cite{schlegel2022comparison}.

\subsection{Item Memory}
It is well known that one of the primary challenges encountered in DNNs revolves around memory consumption. 
The same cannot be said for computational structures based on HV operations, since their overall architecture revolves around a different paradigm that does not include the classic access memory. 
The Item Memory (iM) is a key element in HDC architecture.
It is used as the storage for base HVs that represent the base symbols linked to a classification problem \cite{karunaratne2020memory}.
However, using the same method for representing integers and continuous values in a specific sequence is not suitable.
Therefore, a continuous item memory (CiM) is used where two close numbers have a small distance in their corresponding HD vectors \cite{rahimi2016hyperdimensional}.
Such continuous mapping better represents the adjacent levels since their corresponding HVs are similar to each other.
These concepts are linked with Associative Memory (AM), a computational block that stores all trained HD vectors to be used later for inference. 
The essential function of AM is to compare the incoming encoded query HD vector with the stored classes and return the closest class using the appropriate similarity metrics.

\subsection{HDC Model Deployment}
A system diagram for a general classification task using HDC is presented in Fig.~\ref{fig:general_hdc_scheme}. 
In HDC, the training process is distinct from traditional DL as it does not involve any optimization procedure and it is done in a `one-shot' manner. 
As described in Sect.~\ref{sect:HV_operations}, HDC relies on deterministic vector space operations such as addition, dot product, and permutation, for encoding and processing information. 
This approach contrasts with DL's iterative and computationally intensive weight optimization in multi-layered neural networks.
HDC uses an iM to store and manipulate vectors representing data items or concepts, focusing on vector access and combination rather than function optimization to fit data.
A general learning phase for a classification task in HDC can be described as follows:

\vspace{6pt}
\noindent\textbf{Training Phase:}
\begin{enumerate}
    \item \textit{Encoding Training Data into HD Space:} Training data are mapped into the HD space using randomly generated HVs stored in the iM.
    
    \item \textit{Creating Class Prototypes:} Class prototypes, which correspond to the different classes involved, are generated and stored in the AM, with one prototype for each class. 
    These prototypes are created by using similarity metrics to compare all the HVs associated with a particular class: the HVs that exhibit the highest distance, meaning those with the lowest similarity scores, are then bundled together to form the class prototype.
\end{enumerate}
\vspace{6pt}
\noindent\textbf{Testing Phase:}
\begin{enumerate}
    \item \textit{Query HV Generation:} A query HV $Q$, representing the test data with an unassigned label, is created.
    
    \item \textit{Similarity Assessment in AM:} The query HV $Q$ is compared with stored class prototypes in the AM. Similarity between $Q$ and each class prototype is measured.
    
    \item \textit{Class Label Assignment:} The class prototype with the highest similarity to $Q$ is selected. The label of this prototype is assigned to $Q$, classifying the unknown pattern.
\end{enumerate}

\section{Proposed Methodology}
\label{sect:proposed_method}
\begin{figure*}[!ht]
    \centering
    \includegraphics[width=\textwidth]{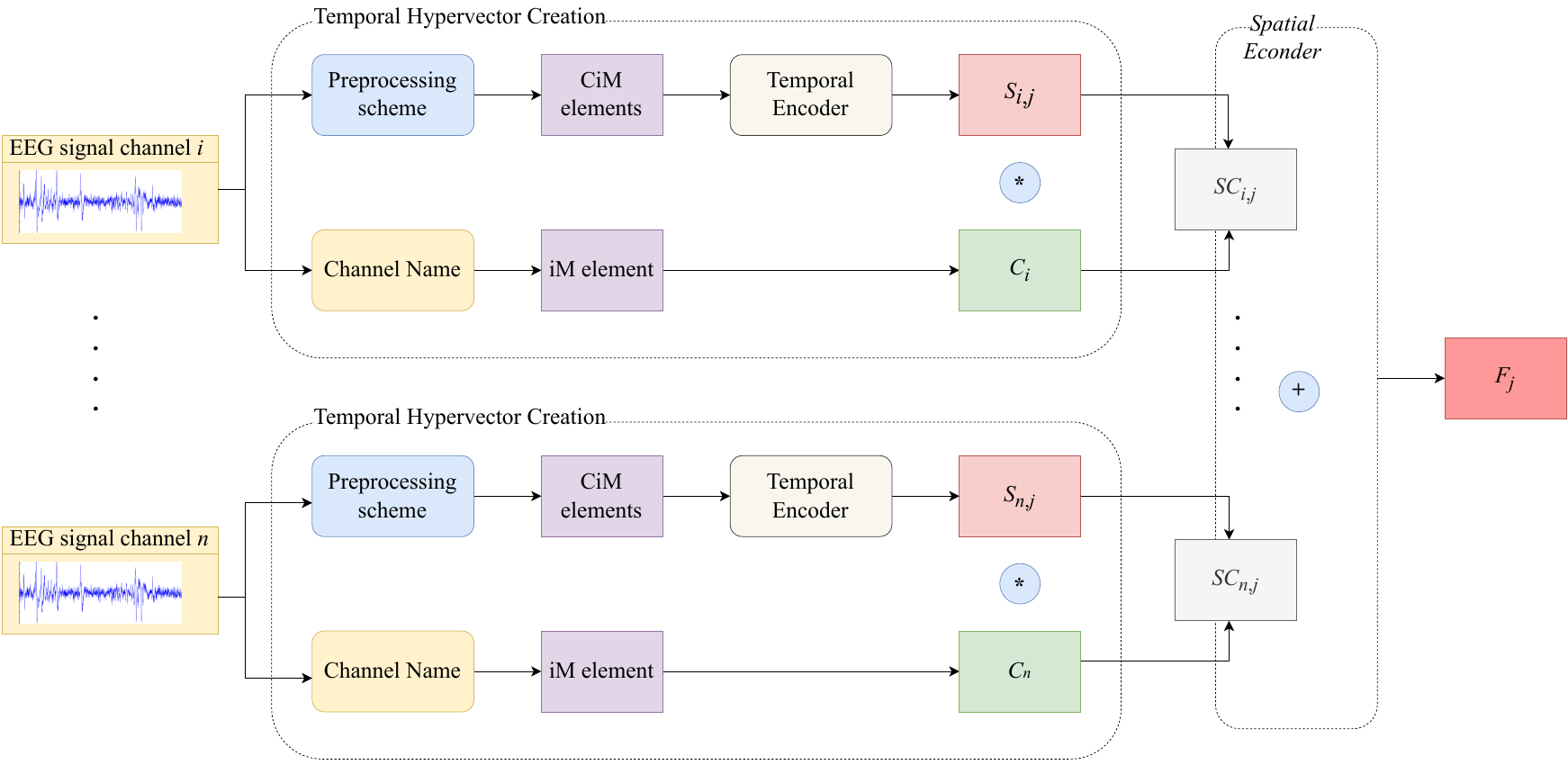}
    \caption{Graphical representation of the spatio-temporal encoder applied to the $j$-th patient. This encoder processes EEG signals, transforming them into an HV representation to be used for the final classification.}
    \label{fig:spatio_temporal_encoder}
\end{figure*}

In this section, it is delineated the use of HDC for ADHD detection: the architecture of the proposed methodology is explained in Fig.~\ref{fig:spatio_temporal_encoder} and we will refer to it as `ADHDC'. 

The standard practice regarding EEG is recording electrical activity (\si{\micro\volt}) using multiple electrodes placed on the scalp. 
Each electrode captures the brain's electrical activity from a specific area of the brain, commonly referred to as `channels'. 

The baseline encoder \cite{rahimi2017hyperdimensional} employs a patient-specific training and testing approach where each patient, with recorded EEG data, is treated as a distinct entity, leading to the initialization of individualized iM and CiM. 
This approach, while potentially offering high degrees of customization and specificity, may not capture the broader generalizability required for the different populations. 
Furthermore, the authors of the paper employed a series of preliminary manipulations on the dataset, which have been omitted in the proposed framework, as described in Sect.~\ref{subsect:novelty}. 

In our solution, the scope is expanded to encompass generalization across entire datasets: instead of treating each patient as a distinct entity, the spatio-temporal encoder is trained using a methodology that partitions the dataset into training and test sets comprising all patients. 
This strategy is designed to create a generalized application of the model, reducing potential biases and constraints associated with patient-specific training procedures. 
The objective is to reduce the number of preprocessing steps involved, allowing the encoder to handle EEG signals that are closer to their original form while still undergoing minimal classic signal processing techniques. 
This approach helps to preserve the essential characteristics of the EEG signals, which could be crucial for identifying particular diagnostic patterns.
Additionally, all encoder parameters are revised and optimized to more closely align with the specifics of the problem, thereby enhancing the model's performance on this specific task.

\subsection{General Overview Of the Spatio-Temporal Encoder}
\label{sect:encoder_overview}
In order to address the temporal nature of EEG data, a temporal encoder is crucial to encompass all signal levels throughout the entire event, similar to the role of recurrent neural networks (RNNs) in traditional DNN architectures. 
However, unlike RNNs, the temporal encoder used herein efficiently captures the entire temporal spectrum of EEG signals in a single encoding step, avoiding challenges like vanishing gradients associated with sequential processing in RNNs.
As already explained, the encoder employed in this paper builds upon the one established in \cite{rahimi2017hyperdimensional} and consists of two components: a spatial encoder and a temporal encoder. 

The spatial encoder is designed to link the electrodes of the same patient by assigning each EEG channel a unique HV, this assignment allows the model to capture the spatial distribution of brain activity, reflecting how specific brain regions contribute to the overall EEG signal. 
Each channel’s HV is drawn from an orthogonal set, ensuring that the spatial information of each channel is distinct, then, through the binding of HVs from different EEG channels, the model can combine spatial information into a single composite HV representation. 
This binding operation enables the model to detect spatially distributed patterns of brain activity that are indicative of ADHD, capturing the interaction between different brain regions (meaning different channels). 
This spatial encoding process not only captures localized brain activity but also integrates it into a broader representation, allowing the model to differentiate between region-specific patterns and more widespread signals, further enhancing its classification capability.

The temporal encoder, on the other hand, transforms the temporal signal levels into a HD representation, preserving the sequence of events across time through permutation operations: EEG signals are divided into non-overlapping $n$-grams, each of which represents a fixed time window. 
The use of $n$-grams allows the encoder to capture local temporal dependencies within the signal, to preserve the order of the temporal information, the permutation operation is applied to these $n$-grams, creating a sequence of permuted HVs that ensure the temporal structure of the EEG data is encoded into the high-dimensional space. 
Permutation acts as a position-dependent transformation, ensuring that the time sequence is embedded into the model's final representation, distinguishing it from time-independent approaches. This method helps capturing time-dependent variations in brain activity, making the model sensitive to both short-term and long-term changes in the EEG signals.

The advantage of this spatio-temporal encoding approach lies in its ability to leverage the concentration of measure phenomenon in high-dimensional spaces: by distributing information across many dimensions, the model becomes robust to small variations in both spatial and temporal data, allowing it to generalize better from fewer samples. 
The final HV encapsulates both spatial and temporal features, which are then compared against class prototypes stored in the AM using similarity metrics.

\subsection{$n$-grams Creation and Spatio-Temporal Encoder}
As noted in \cite{kanerva2009hyperdimensional}, $n$-grams are a fundamental tool in HDC, offering a method for encoding sequential information into high-dimensional spaces.
Based on this, the first step of the HDC encoder, explained in Sect.~\ref{sect:encoder_overview}, involves partitioning EEG sequences into non-overlapping sets of $n$-grams, each one with a fixed dimension of $32$. 
The decision to use this specific sample size was strategically made to encapsulate one second of EEG data, originally sampled at $256$ Hz and then downsampled to $32$ Hz, within a single $n$-gram. 
This approach ensures that essential information is retained while reducing the data complexity; the validity of this downsampling choice was confirmed by testing the model across different configurations of $n$-grams and downsampling dimensions, aiming at optimized performance while maintaining efficiency in data processing.

From each complete EEG sequence, representing each patient, a total of $28$ $n$-grams $N_{i}$ are derived for each EEG channel, representing the segmented temporal activities of the signal levels of the $i$-th electrode.
Once the $n$-grams are created through simple subsequences extractions from the EEG, the next step utilizes the encoder to generate their HDC representation.

\begin{figure*}[!ht]
    \centering
    \includegraphics[width=0.9\textwidth]{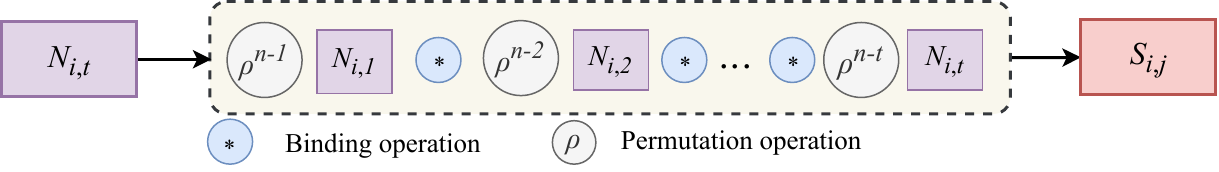}
    \caption{Graphical scheme of the temporal encoder: the EEGs signal levels over time are transformed into a hyperdimensional representation.}
    \label{fig:temporal_encoder}
\end{figure*}

The temporal encoder proposed is summarized in Fig.~\ref{fig:temporal_encoder}: 
the inputs are the sequences of $n$ quantized signal levels from the $i$-th electrode of the patient $j$ and computes the $n$-gram HV $S_{i,j}$. 
It defines the HV representation of the temporal activities of the signal levels for the $i$-th electrode for patient $j$.
Specifically, to encode a sequence of signal levels, a permutation operation is used, denoted as $\rho(\cdot)$, forming an $n$-gram HV.


%
%

A generalization for $n$-grams of different sizes, applied to a patient $j$, can be described as: 
\begin{equation}
S_{i,j} = \prod_{t=1}^{n} \left( \rho^{n-t} \times N_{i,t} \right)\,.
\end{equation}
Lastly, to connect the temporal series to the specific channel, we bind $S_{i,j}$ with its channel name $C_{i}$ (the HVs initialized in the iM) using point-wise multiplication and obtaining in this way $SC_{i,j}$: 
\begin{equation}
SC_{i,j} = S_{i,j} \times C_{i}\,.
\end{equation}
After projecting the EEG signal values into the hyperspace, the multiplication operation is used to bundle each channel $C_{i}$ to its signal level $N_{i,t}$ for every timestamp $t$. 

At the end of this process, we employ the spatial encoder to bind together the two channels for each patient $j$ to have the entire EEG event represented by a single HV $F_{j}$ that in this way can encapsulate both the spatial and temporal dimension of the EEG.
In other words: given $S_{i,j}$, the encoded representation of the quantized signal for the channel $i$ and patient $j$, and $C_{i}$ the iM element corresponding to the channel $i$ we have that the final encoded representation for patient $j$: 
\begin{equation}
F_{j} = \sum_{i=1}^{2} (SC_{i,j})\,.
\end{equation}

\subsection{iM and CiM Initialization}
Given that each electrode possesses a distinct string identifier, it constitutes a field that can be readily correlated with an HV.
Consequently, an equivalent number of HVs $C_{i}$ is created within the iM corresponding to the total count of channels present in the recorded EEGs, ensuring that each item remains orthogonal and independent from every other HV.
Since we're working with biosignal processing applications, an alternative mapping approach to HVs for what concerns the temporal sequence is needed. 
For this reason, the CiM is initialized using the processes described in \cite{rahimi2016hyperdimensional}. 
Firstly, the EEG signal values from each electrode are subjected to a linear scaling process, ranging from $e_\mathrm{min}$ to $e_\mathrm{max}$, and subsequently quantized into $e_\mathrm{max}$ distinct levels. Based on different trials, we opted for $e_\mathrm{min}$ set to $0$ and $e_\mathrm{max}$ set to $250$.
Then, to encode the $e_\mathrm{max}$ discrete signal levels, the CiM maps these levels onto HD vectors: for the lowest signal level $e_\mathrm{min}$, the CiM designates a unique orthogonal HV $CN_{i,t}$, while the remaining $e_\mathrm{max}$ levels are then sequentially generated, each one progressively diverging from the HV of the minimum level. 
This progression is designed such that the HV for the highest signal level $e_\mathrm{max}$ is orthogonal to the one in the lowest level. 
Orthogonality in this context means that two HVs differ in half of their components. 
This approach of gradual differentiation ensures that HVs of adjacent signal levels are similar, allowing for a more continuous and representative mapping of signal values.

\subsection{Training and Testing}
As described in Sect.~\ref{sect:hdc_fundamentals}, the training process in HDC relies solely on intrinsic operations of HD computing without involving any optimization procedures.
The process began by dividing the dataset into a training set and a test set, assigning $59$ patients to the training set, with $27$ patients from the ADHD group and $32$ from the Control group. 
For the test set, we allocated $20$ patients, ensuring an equal division between ADHD and Control groups. 
The procedure for the training process is summarized in the pseudo-code of Algorithm~\ref{alg:train_scheme}.

\begin{algorithm}
\vspace{6pt}
\caption{Training Phase of the Spatio-Temporal Encoder}
\label{alg:train_scheme}
\begin{algorithmic}
    \item[1.] \textbf{Initialize:}
    \item[1.1] Set dimension $D = 10000$
    \item[1.2] Generate bipolar dense HVs with $D$ dimensions, values randomly $+1$ or $-1$
    
    \item[2.] \textbf{Create Class Prototypes:}
    \item[2.1] Initialize $HV_A$ (ADHD class) and $HV_C$ (Control class) in memory as empty vectors with $D$ dimensions.
    
    \item[3.] \textbf{Training Process:}
    \item[3.1] For each patient $j$, generate HV $F_j$ using spatio-temporal encoding for patient $j$
    \item[3.2] For each class prototype in \{$HV_A$, $HV_C$\}:
    \begin{enumerate}
        \item[3.2.1] Compute cosine similarity between $F_j$ and the prototype
        \item[3.2.2] If cosine similarity $< 0.5$, bundle $F_j$ with the prototype
    \end{enumerate}
\end{algorithmic}
\end{algorithm}

In this way, during testing the class prototypes are iteratively updated to incorporate new, sufficiently distinct patterns from the EEG data.
Then, after establishing the Class Prototypes, it is possible to proceed to the testing phase following algorithm presented in the pseudo-code of Algorithm~\ref{alg:test_scheme}.

\begin{algorithm}
\vspace{6pt}
\caption{Testing Phase of the Spatio-Temporal Encoder}
\label{alg:test_scheme}
\begin{algorithmic}
    \item[1.] \textbf{Creation of the Query HV \(Q\):}
    \item[1.1] Assign the output of the spatio-temporal encoder from the test set as a query HV \(Q\)

    \item[2.] \textbf{Forwarding \(Q\) to the Associative Memory (AM):}
    \item[2.1] Send the \(Q\) HV to the AM to determine its originating class

    \item[3.] \textbf{Classification Process:}
    \item[3.1] Initialize \texttt{similarity\_A} as the cosine similarity between \(Q\) and \(HV_A\)
    \item[3.2] Initialize \texttt{similarity\_C} as the cosine similarity between \(Q\) and \(HV_C\)

    \item[4.] \textbf{Final Classification:}
    \begin{enumerate}
        \item[4.1] If \texttt{similarity\_A} > \texttt{similarity\_C}, designate the originating class of \(Q\) as ADHD (Class A)
        \item[4.2] Else, designate the originating class of \(Q\) as CONTROL (Class C)
    \end{enumerate}
\end{algorithmic}
\end{algorithm}

\section{Experimental Results}
\label{sect:experiments}
In this section, we present experimental results and sensitivity analysis of our ADHDC model, focusing on evaluating its classification accuracy and determining the minimum number of patients required to achieve benchmark accuracy.

\subsection{Experimental Setup}
The dataset used in our work was released in April $2023$ and comprises EEG signals of $37$ ADHD and $42$ non-ADHD patients \cite{datasetADHD}. 
EEG is a non-invasive technique used to record electrical activity in the brain. 
EEG measures and records the electrical impulses generated by neurons when they communicate with each other.
This technique involves placing electrodes on the scalp which detect and amplify these electrical signals. 
These electrodes pick up the brain's electrical activity and produce a graphical representation known as an EEG recording or EEG signal. 
EEG is commonly used in neuroscience, clinical medicine, and research to study brain function, diagnose different neurological conditions, monitor sleep patterns, and assess brain activity in response to stimuli or tasks \cite{cohen2017does}.

In this study, EEG data were recorded from patients in four states, based on their activity: resting with eyes open, eyes closed, during a cognitive challenge, and while listening to omni-harmonic stimuli \cite{datasetADHD}. 
The EEGs, originally sampled at $256$ Hz, were collected from five channels (O1, F3, F4, Cz, Fz), each representing a different brain area. 
However, for each task, only two specific channels were used, based on the brain area of interest. 
Our study focused on the baseline task with eyes open, utilizing data from channels F4 and Cz which were selected to avoid contaminating the EEG readings with the influence of various tasks being conducted.

\begin{figure*}[!ht]
    \centering
    \includegraphics[width=0.95\textwidth]{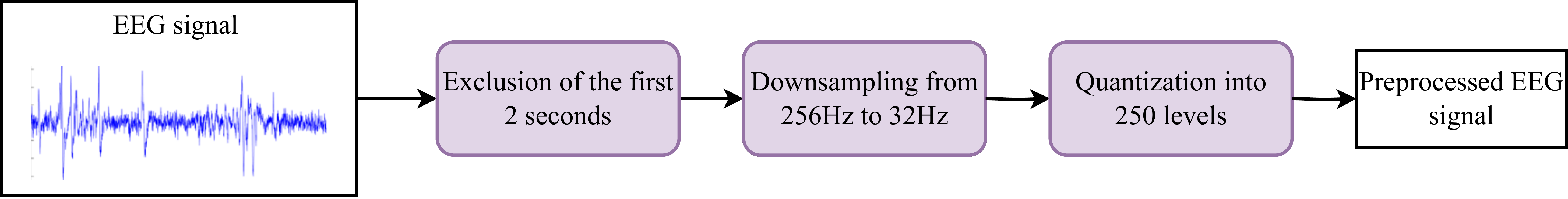}
    \caption{Preprocessing steps applied to each EEG signal.}
    \label{fig:preprocessing_eeg}
\end{figure*}

As already explained in Sect.~\ref{subsect:novelty}, one of the primary objectives of our study was to limit data preprocessing to the bare minimum: only the essential and non-medical related preprocessing steps to the EEG data have been applied. 
These steps are concisely summarized and illustrated in Fig.~\ref{fig:preprocessing_eeg}. 

\begin{figure}[!ht]
    \centering
    \includegraphics[width=0.8\columnwidth]{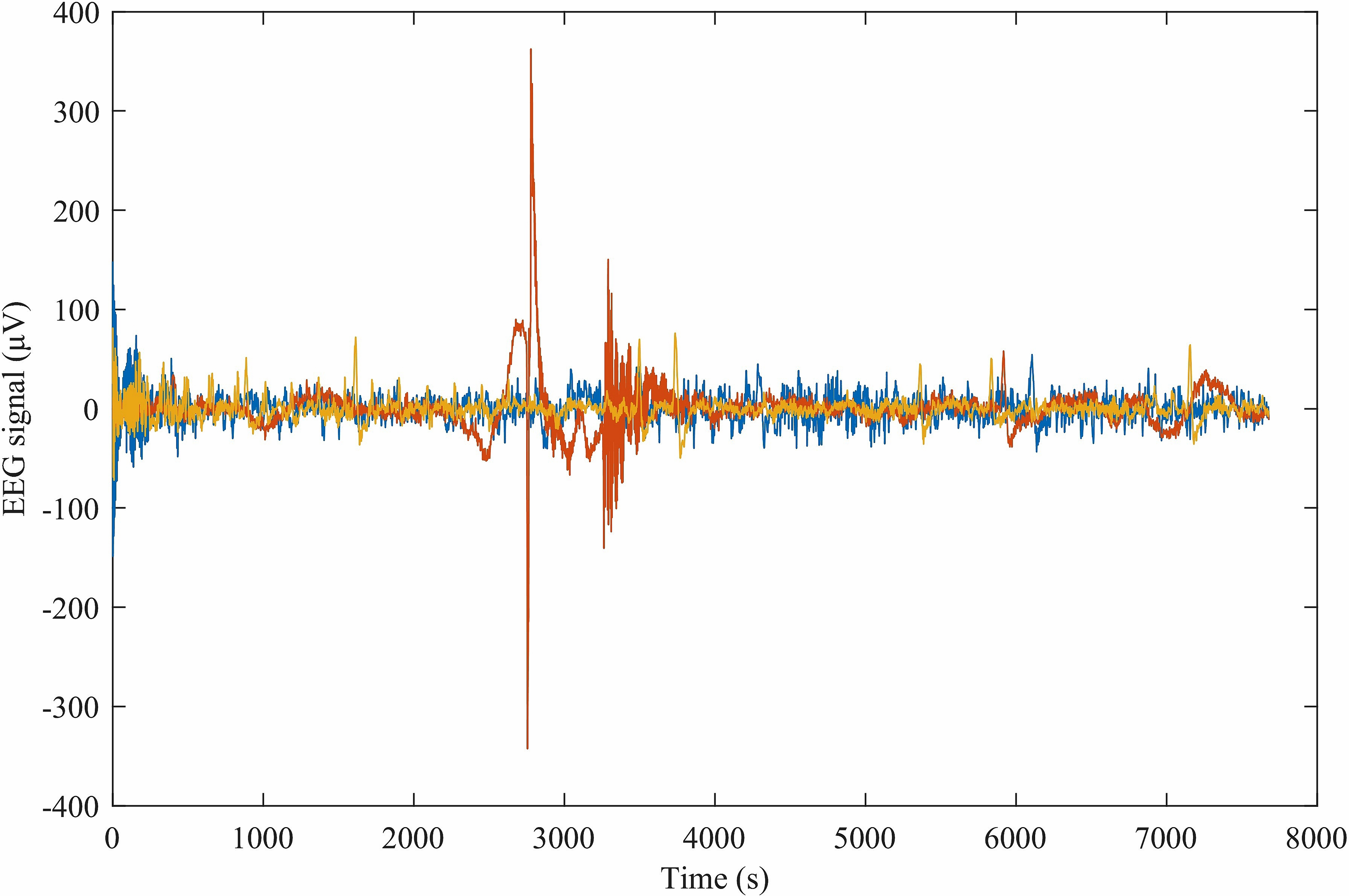}
    \caption{Graphical representation of key EEG signals, showcasing a selection of the most illustrative and characteristic EEGs recordings from the dataset.}
    \label{fig:plot_eeg}
\end{figure}

A series of EEG data plots in their raw, unprocessed form is illustrated in Fig.~\ref{fig:plot_eeg}.
The visualization shows a sparse distribution of values above the set threshold (red signal), with noticeable external noise in the initial segment (blue and orange signals).
Each EEG was recorded for $30$ seconds with a sampling rate of $256$ Hz, ending up with an initial shape of $(v_\mathrm{raw},ch)$ where $v_\mathrm{raw} = 7680$ represents the number of temporal sequence values for each of the $ch$ channels, where $ch = 2$ indicates the total number of channels employed. 

We chose to exclude the initial $512$ signal samples, corresponding to the first two seconds of the EEG, as these are typically the noisiest due to the settling phase of the examination \cite{gorjan2022removal}, resulting in a total of $7168$ signal values for analysis.
Furthermore, to carry out the initial quantization step, it was necessary to determine the minimum and maximum values for each channel across the entire dataset. 
To preserve the rawness of EEG signals, while handling outliers, we set upper and lower thresholds in the dataset, addressing anomalies without heavily altering the data. 
Lastly, we downsampled the EEG frequency from $256$ Hz to $32$ Hz by averaging every $8$ values, reducing data points for computational efficiency, and preserving essential EEG details while optimizing them.
This operation led to the final shape of the dataset: ($p$, $v_\mathrm{final}$, $ch$) where $p = 79$ is the number of patients, $v_\mathrm{final} = 896$ is the number of values related to the EEG and $ch = 2$ is the number of channels. 

\subsection{Numerical Results}
\label{subsect:num_results}
\begin{figure}[!ht]
    \centering
    \includegraphics[width=0.7\columnwidth]{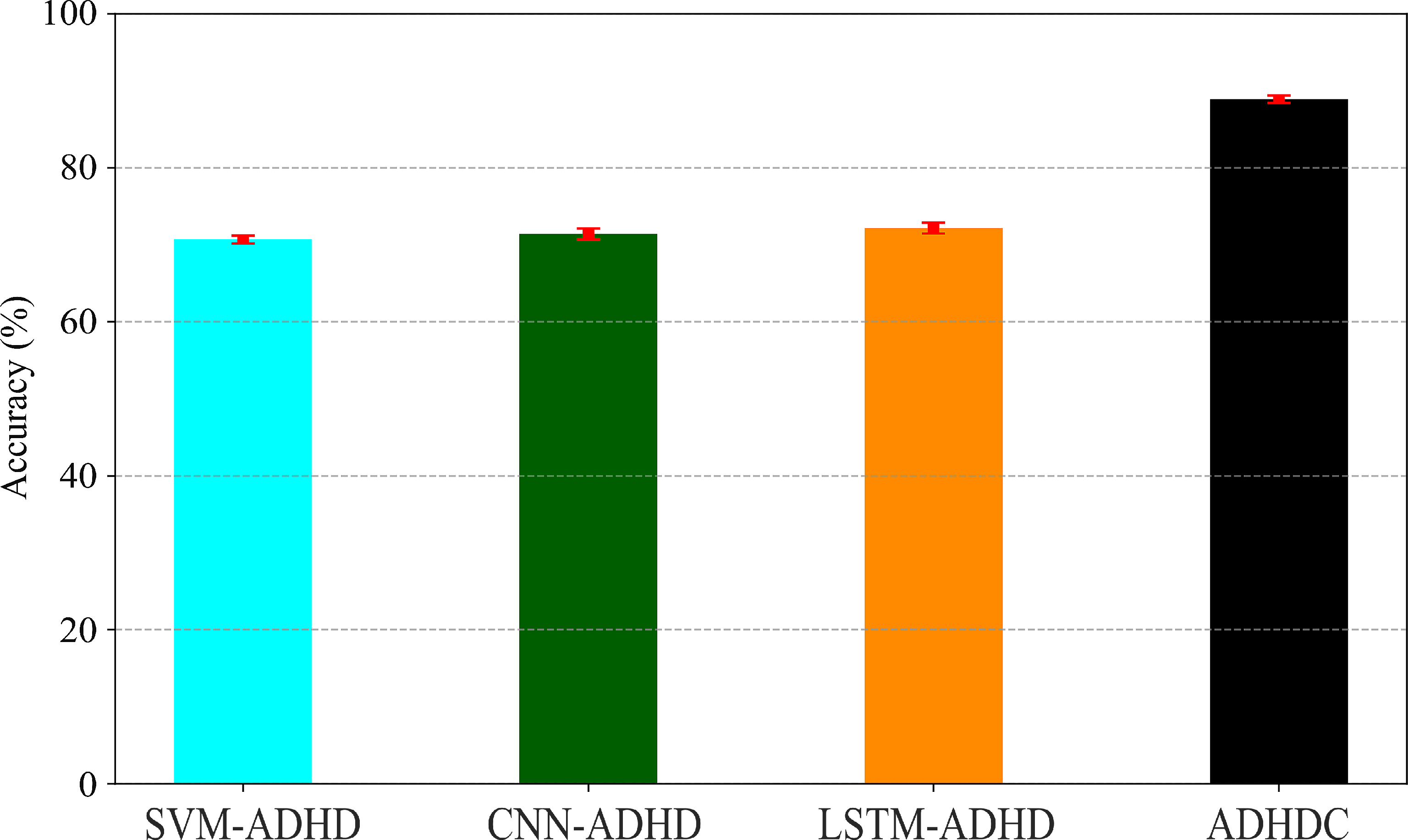}
    \caption{Bar Plot of the average accuracy for the three models on the test set.}
    \label{fig:bar_plot_classification}
\end{figure}

All the experiments were conducted using MATLAB \textsuperscript{\textregistered} back-end on a machine equipped with an Intel{\textsuperscript{\textregistered}}  Core{\textsuperscript{\texttrademark}} i$5$-$10210$U CPU at $2.11$ GHz, $8$ GB of RAM. 
Results were obtained as an average over $r$ different runs, where $r = 10$, of the model, where each run was characterized by a different random partition among training and test sets. 
The average accuracy reached by our implemented model is described in Fig.~\ref{fig:bar_plot_classification}, together with the accuracy performance of the benchmarking models. 

To evaluate the effectiveness of this not yet utilized dataset, three benchmark models are created and employed: a Support Vector Machine (named `SVM-ADHD'), a Convolutional Neural Network (named `CNN-ADHD') and lastly a Long Short-Term Memory network (named `LSTM-ADHD').
SVM-ADHD is adapted from \cite{hsu2003practical} and employs an RBF kernel and hinge loss function, CNN-ADHD is a basic CNN with two Convolutional layers and $3$ Fully Connected and LSTM-ADHD has $2$ LSTM layers and $3$ Fully Connected. 
The hyperparameters of all three networks have been fine-tuned for each model through a grid search procedure to identify the optimal configuration. 
The final test accuracy, along with its standard deviation and F1-score, Precision, Recall and Sensitivity metrics of each model is summarized in Table~\ref{tab:bench_acc} and shown through the bar plot in Fig.~\ref{fig:bar_plot_classification}.

\begin{table*}[!ht]
\centering 
\caption{Numerical results of the accuracy on the test set for the benchmark models and the presented one (in bold), averaged over 5 different runs.}
\label{tab:bench_acc}
\begin{tabular}{lcccc}
\toprule
\textbf{Model} & \textbf{Accuracy (\%)} & \textbf{F1-Score} & \textbf{Precision} & \textbf{Recall} \\
\midrule
SVM-ADHD & $70.7 \pm 0.05$ & $0.712 \pm 0.003$ & $0.708 \pm 0.004$ & $0.716 \pm 0.002$ \\ 

CNN-ADHD & $71.4 \pm 0.01$ & $0.720 \pm 0.002$ & $0.715 \pm 0.003$ & $0.727 \pm 0.001$ \\ 

LSTM-ADHD & $72.2 \pm 0.01$ & $0.731 \pm 0.003$ & $0.729 \pm 0.002$ & $0.733 \pm 0.003$ \\

\textbf{ADHDC} & $\bf{88.9 \pm 0.02}$ & $\bf{0.875 \pm 0.004}$ & $\bf{0.850 \pm 0.005}$ & $\bf{0.904 \pm 0.003}$ \\
\bottomrule
\end{tabular}
\end{table*}

Given our approach of segmenting patients' EEGs into smaller $n$-grams and deriving a set for each patient, coupled with our decision to classify individual $n$-grams, we have established a criterion for determining the overall correctness of a patient's classification. 
As explained in Sect.~\ref{sect:proposed_method}, patients' EEGs are segmented into $n$-grams and classified individually and we deem a patient as correctly classified if the number of correctly classified $n$-grams is strictly greater than half of the total number of $n$-grams, serving as the threshold for classification accuracy. 
Once evaluated if a patient is correctly classified or not, the final performance is measured on the test set by the standard classification accuracy:
\begin{equation}
A = \frac{100}{\text{{S}}} \sum_{i=1}^{S} |t_i - y_i|\,,
\end{equation}
where $S$ is the number of total patients to be tested; $t_{i} \in \{0,1\}$ is the target binary label of the $i$-th sample, being $t_{i} = 0$ the label of a neurotypical child and $t_{i} = 1$ the one of a child with ADHD; $y_{i} \in \{0,1\}$ is the binary label estimated by the classifier on the same sample. 

Our HDC classifier exhibits an average accuracy of $88.9\%$ which is $23.12\%$ higher than the baseline with the same conditions.
Besides, only $9$ out of $79$ patients are misclassified, with $6$ being false positives, further highlighting the precision of the HDC approach.
The relatively low number of misclassifications indicates a high level of sensitivity and specificity, essential qualities in medical diagnostic tools.
Furthermore, leveraging HDC's acknowledged capability to achieve high accuracies with limited training data, we conducted an empirical investigation to identify the minimum number of patients needed to surpass the LSTM-ADHD accuracy of $72.2\%$. 

The proposed approach involved incrementally training the model with varying patient sample sizes, starting with a single patient, and consistently evaluating its performance on a fixed test set of $20$ patients. 
The iterative process involved increasing the number of patients in the training set one at a time, up to $59$ patients (as the total dataset comprised $79$ patients with $20$ fixed for testing) while training, and testing it each time on the same test set. 
Each training iteration was consistently evaluated using the fixed test set to ascertain the minimum number of patients needed to train the HDC model effectively, aiming to match or exceed the performance of the LSTM-ADHD model. 

\begin{figure}[!ht]
    \centering
    \includegraphics[width=.7\columnwidth]{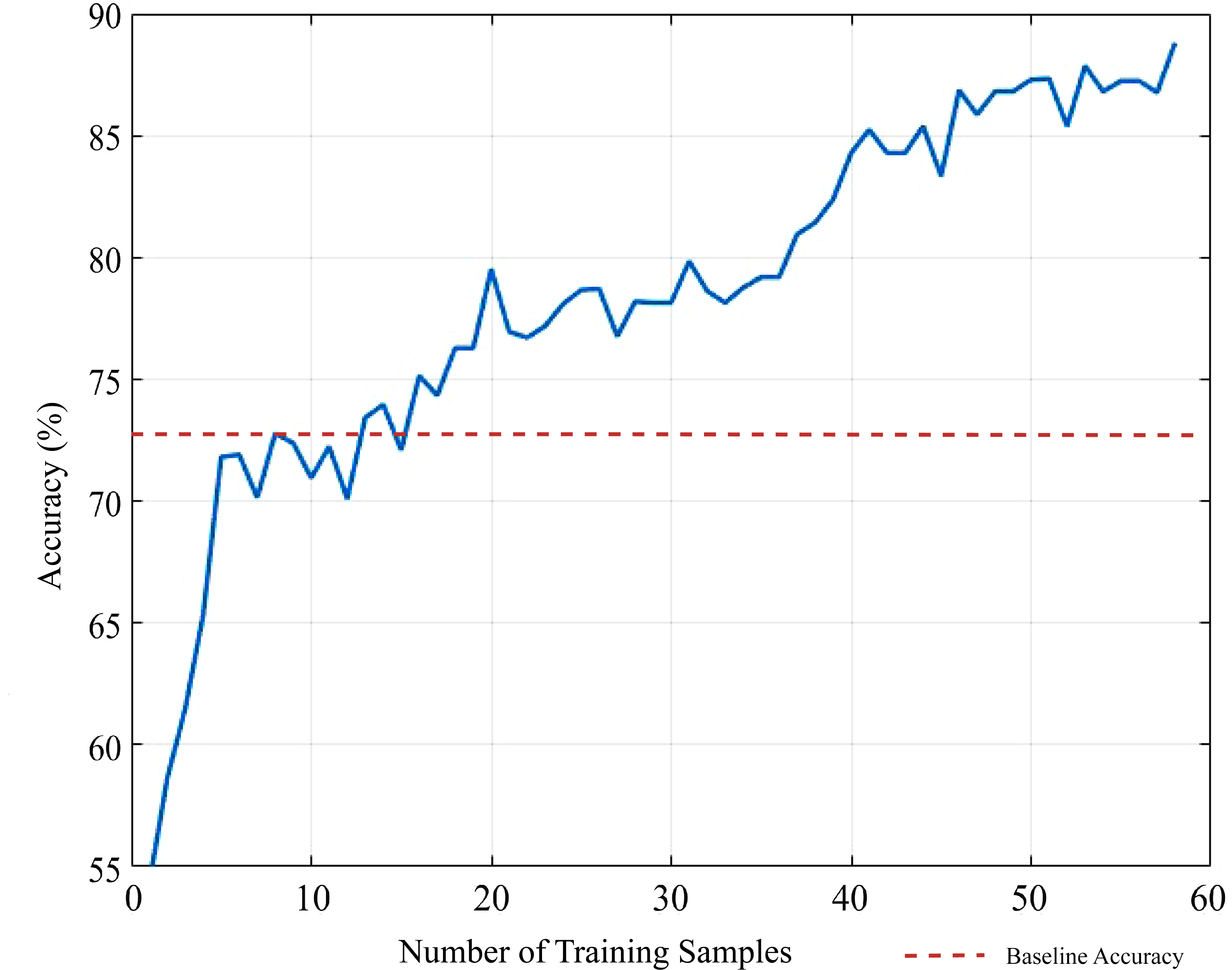}
    \caption{Accuracy on test set given the incremental number of training patients.}
    \label{fig:acc_incremental}
\end{figure}

Moreover, as illustrated in Fig.~\ref{fig:acc_incremental}, the average accuracy of the HDC classifier exhibits notable growth with the increasing number of training patients. The HDC model's ability to generalize from small data samples and understand underlying patterns, rather than just memorizing data, underscores its effectiveness in data utilization, making it a practical tool where accuracy and computational efficiency are crucial.
%
%


\subsection{Discussion}
\label{sect:discussion}
The numerical results presented in Sect.~\ref{subsect:num_results} demonstrate that our HDC framework for ADHD classification achieves strong performances, with an accuracy of up to $88.9\%$. Notably, when trained on just $8.86\%$ of the dataset, the model reached an accuracy of $72.8\%$, surpassing the $72.2\%$ achieved by the LSTM-ADHD model trained on the full dataset. This highlights data efficiency of the HDC model, a significant advantage in medical applications where large datasets are often difficult to collect, this aspect is particularly relevant for detecting neurodevelopmental disorders such as ADHD, where medical data is often scarce, especially in emerging research areas involving evidences from objective data sources as neuroimaging.

Unlike DNNs, which typically require large amounts of data to generalize effectively, HDC provides a robust solution for scenarios with limited data: the ability to achieve strong generalization with small datasets makes HDC a promising tool for medical diagnostics, particularly in fields like ADHD, which are still in the early stages of research. 
One key factor behind HDC’s superior performance is its high-dimensional representation, where information is distributed across many components of a HV. This distributed encoding allows the model to capture essential patterns even with a small number of training samples.

The efficiency of the HDC model is further demonstrated by its minimal computational requirements, achieved through lightweight operations that eliminate the need for GPUs or high-performance hardware. This is evidenced by its rapid training time of just $15$ seconds and an inference time of only $0.001$ seconds for a single $n$-gram classification. These results were obtained on the standard, consumer-grade device described in Section \ref{subsect:num_results}, highlighting the model’s ability to deliver high performance even on non-specialized hardware.
    These characteristics highlight its suitability for real-time applications, such as ADHD screening in clinical or field settings, where low latency and high throughput are essential. In contrast, traditional models such as CNN-ADHD and LSTM-ADHD require more than $4$ minutes to train on the same hardware and have significantly higher computational overhead. The combination of rapid inference and low computational complexity underscores the practicality of HDC for deployment in resource-constrained environments.

Additionally, ADHDC achieved a F1-score of $0.875$, substantially outperforming other models and indicating a balanced level of precision and recall. 
The model's high recall of $0.904$ suggests that HDC effectively minimizes false negatives, which is crucial in clinical diagnostics: it’s noteworthy that the model produces more false positives than false negatives, which is beneficial in this context. 
Since objective diagnostic tools like HDC are intended to complement, not replace, traditional diagnostic methods, having more false positives leads to additional follow-up tests rather than risking missed diagnoses. This characteristic makes HDC a valuable support tool, as any false positive can prompt further examination, ultimately contributing to a more accurate and thorough diagnostic process.

\section{Conclusions}
\label{sect:conclusions}
This study introduced a novel application of HDC for ADHD classification using EEG signals: the proposed framework achieved an accuracy of $88.9\%$, significantly outperforming traditional ML models such as SVM, CNN, and LSTM, as shown in Sect.~\ref{subsect:num_results}. Notably, the HDC framework demonstrated an exceptional capacity to deliver high performance with limited data, achieving an accuracy of $72.9\%$ using only $8.86\%$ of the dataset, that corresponded to just seven patients. This result surpasses the performance of the highest benchmark, LSTM-ADHD, trained on the full dataset, emphasizing HDC's efficiency and robustness in handling small sample sizes, a common limitation in medical and clinical settings. Additionally, the proposed model required minimal data preprocessing and employed EEG data with minimal preprocessing, making the approach both scalable and practical for real-world clinical applications.

For future work, several promising directions could be explored. First, enhancing the interpretability of the learned features would be valuable: investigating how these features correspond to ADHD-specific EEG patterns could significantly improve the model’s clinical utility. Conducting a detailed analysis of attention maps to highlight the most relevant EEG features for classification would make the model’s predictions more transparent and easier for clinicians to understand, offering insights into its decision-making process and supporting the validation of results with evidences from clinical studies.
Furthermore, integrating HDC with neural networks could combine the strengths of both approaches, merging the pattern recognition capabilities of DL with the efficiency and scalability of HDC: this hybrid model could be developed further for tasks requiring both high accuracy and computational efficiency. Expanding the framework to include other neurodevelopmental disorders would also broaden its application and impact in medical diagnostics. 
Finally, optimizing the model for deployment on mobile or embedded platforms would enable real-time ADHD screening in resource-constrained environments, making the technology more accessible in clinical settings.

\bibliographystyle{IEEEtran}
\bibliography{IEEEabrv,ref}


\begin{IEEEbiography}[{\includegraphics[width=1in,height=1.25in,clip,keepaspectratio]{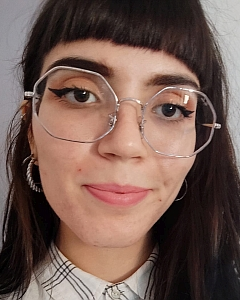}}]{Federica Colonnese} was born in in 1997. She received the M.Sc. Degree in Industrial Engineering (Business Intelligence \& Analytics curriculum) from University of Rome ``La Sapienza'', Italy. She is currently a Ph.D. Student in Information and Communication Technology from the same University and her main interests concern the development of deep learning models for the detection of neurodevolpmental disorders.
\end{IEEEbiography}

\begin{IEEEbiography}[{\includegraphics[width=1in,height=1.25in,clip,keepaspectratio]{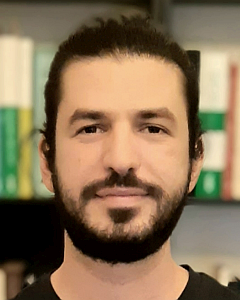}}]{Antonello Rosato} (S’17–M’19) was born in 1990. He received the Ph.D. in Information and Communication Technology from the University of Rome ``La Sapienza'', Italy, where he is currently an Assistant Professor at the Department of Information Engineering, Electronics and Telecommunications. His research interests include machine learning techniques for prediction of complex behaviors, neural and fuzzy-neural models, and distributed clustering algorithms.
\end{IEEEbiography}

\begin{IEEEbiography}[{\includegraphics[width=1in,height=1.25in,clip,keepaspectratio]{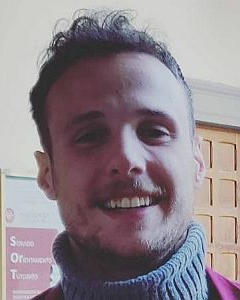}}]{Francesco Di Luzio} was born in 1996. He received the Ph.D. in Information and Communication Technology from the University of Rome ``La Sapienza'', Italy, where he is currently a Research Fellow on topics relating to deep learning and explainable AI applied to emotion recognition and behavioral analysis, time series prediction in the renewable energy sources field, and quantum computing for fraud detection.
\end{IEEEbiography}

\begin{IEEEbiography}[{\includegraphics[width=1in,height=1.25in,clip,keepaspectratio]{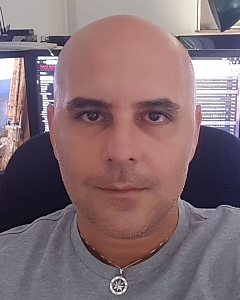}}]{Massimo Panella} (S’01–M’02–SM’16) was born in 1971. In 2002, he received the Ph.D. in Information and Communication Engineering from the University of Rome ``La Sapienza'', Italy. He is currently a Full Professor of electrical engineering, quantum computing and applied machine learning with research interests on the modeling, optimization, and control of real-world systems and for solving both supervised and unsupervised learning problems on real data.
\end{IEEEbiography}

\vfill

\end{document}